\documentclass[aps,epsfig,multicol,amstex]{revtex4}
\usepackage{graphicx}
 \usepackage{amsmath}
 \usepackage{color}
\begin{document}

\title{Cell theory for glass-forming materials and  jamming matter, combining free volume and cooperative rearranging regions}

\author{ Antonio Coniglio$^1$ and Tomaso Aste$^2$}
\address{ $^1$ CNR-SPIN, Dipartimento di Fisica, Universit\`a ``Federico II'', Napoli, Via Cintia, 80126 Napoli, Italy.}
\address{$^2$ Department of Computer Science, University College London, Gower Street, London, WC1E 6BT, UK }
\address{e-mail:  t.aste@ucl.ac.uk }
\date{\today}

\begin{abstract}
We investigate the statistical mechanics of glass-forming materials and jamming matter by means of a geometrically driven approach based on a revised cell theory.
By considering the system as constituted of jammed blocks of increasing sizes, we obtain a unified picture that describes accurately the whole  process from low densities to limit densities at the glass/jamming  transition.
The approach retrieves many of the aspects of  existing theories unifying them into a coherent framework.
In particular, at low densities we find a free volume regime, based on local relaxation process, at intermediate densities a cooperative length sets in, where both local and cooperative relaxation process are present. At even higher densities the increasing cooperative length suppresses the local relaxation  and only the cooperative relaxation survives characterized by the divergence of the  cooperative length, as suggested by the random first order theory. Finally a relation between the cooperative length and the hyperuniform length
is also suggested.

\end{abstract}

\maketitle

There are many systems, such as molecular liquids, colloids,
granular materials, foams and others that by changing the control
parameters exhibit a very slow dynamics followed by a structural
arrest or quasi structural arrest.  Although many progresses have
been made, understanding the glass transition, or jamming,  from a
fluid phase to an amorphous solid is still one of the major problems
in condensed matter.  In particular, the quest concerns understanding the main mechanism for the
huge increase of the viscosity, or relaxation time, as the
temperature decreases or the density increases that is observed across different systems.

Many theories have been proposed. Among the first ones is the Free
volume theory mostly developed by Cohen and Turnbul \cite{cohen1961,cohen1959,cohen1969}, and the
Cooperative rearranging regions (CRR) of Adam and Gibbs \cite{AG}. 
Both, using simple and physical intuitive arguments, predict  a strong exponential
divergence at the ideal glass transition characterized by a
Voghel-Fulcher-Tamman (VFT) law. Free volume theory is based on a local
property of the particles: each particle  can move in a free volume
where the center of a particle can translate, given that all
other particles are fixed. Roughly we can say that the particle
rattles in a cage of its free volume. At very high density, or at low
temperature, close to the glass transition, relaxation occur when
very rarely a particle ends up in a cage with large enough free
volume and manage to jump out of the cage. A somehow opposite
view   was assumed by Adam and Gibbs, who argued that relaxation
occurs due to a cooperative rearrangement of group of particles.
The linear size of such region $\xi$ diverges at a temperature
$T_0$, the Kautzmann temperature \cite{kautz}, where the configurational entropy
vanishes.

An important advance in the theory of the glass transition is
represented by the Mode coupling theory mostly developed by G\"oetze
and collaborators \cite{gotze1991liquids,gotze2009}. Starting from first-principles they derived
an equation for the time dependent density autocorrelation
function, which makes a number of precise dynamical predictions. In
particular, they predicted that the relaxation time should diverge as a
power law at a critical temperature, where structural arrest is
reached. However, molecular glasses do not show such power law
divergence and, nowadays, $T_c$ is considered a crossover towards a
hopping regime characterized by exponential type of the relaxation
time. The nature of the approximations involved in the mode
coupling theory, has been elucidated by the spin glass theory
which predicts, in mean field approximation, the same solution, in
its schematic version, as the mode coupling theory.

In recent years Random First Order Transition (RFOT), first introduced by
Kirkpatrick, Thirumalai and  Wolynes \cite{kirkpatrick1989scaling} and later developed by Wolynes and collaborators\cite{xia2000fragilities,lubchenko2007theory}, has become one of the
most popular theory of the glass transition,  which, besides the
idea of the cooperatively rearranging regions, combines also
features of spin glass  \cite{mezard1999thermodynamics} and mode coupling
theory. It predicts an apparent  power law divergence of the
relaxation time, followed by a Voghel-Fulcher-Tamman law diverging at the Kautzmann
temperature. 
Each theory captures some correct
aspect  of the glass transition. It is then plausible that the
"final"  theory must reproduce the essential features of  these
different approaches.

In this paper, 
we show that a revised version of the Cell Theory of the Glass Transition \cite{AsCo04}  reproduces
 the essential ingredients of both  Free Volume and
Cooperative Rearranging Regions in a unified manner leading to new
predictions.

{\bf  Free Volume Theory}.  Let $N(n)$  be the number of particles
with free volume $v_f(n)$, where $n$ is a discrete index. The
free volume distribution  can be calculated by maximizing, $\Omega({N(n)})$, the number
of ways of redistributing the free volume, $
\Omega({N(n))}= N! /{ \prod_{n}N(n)! }
$
under the following
constraints $\sum N(n)v_f(n) = V_f$ and $\sum N(n) = N$, where
$V_f$ is the total free volume and $N$ is the number of particles.

After maximizing $\Omega({N(n)})$ and passing to the continuum limit one
obtains

\begin{equation}\label{pvf}
p(v_f) = \frac{1}{\left< v_f \right>} \exp(-v_f/{\left< v_f \right>})
\;\;\;\;,
\end{equation}
where $p(v_f)dv_f$ is the probability  to find a particle in a
free volume between $v_f$  and $v_f+dv_f$ and ${\left< v_f \right>} =
V_f/N$ is the average free volume.
In this theory the relaxation time $\tau$ is related to the inverse of the probability that
a particle jumps out of the cage.
If a particle is in a cage with small free volume  $v_f$ the
probability to escape $P_{esc} $  will be small. In a simple approach it is
assumed that,  below a threshold $v_f<v_f^*$, the particle is
localized, otherwise it can jump and get out of the cage.
In this case  the probability $P_{esc}$ that a given particle  jumps
out of the cage is given by the probability that the particle has
a free volume $v_f>v_f^*$ :

\begin{equation}\label{tau}
P_{esc} = \int_{v_f^*}^{\infty} p(v_f)\, dv_f
\;\;\;\;,
\end{equation}
where $p(v_f)$ is given by  Eq.\ref{pvf}.
The relaxation time $\tau/\tau_0$ (where $\tau_0$ is a microscopic time) is proportional to the inverse of this probability:

\begin{equation}\label{taufreev}
\frac{\tau}{\tau_0} = \frac{1}{P_{esc}} = \exp(v_f^*/{\left< v_f \right>})
\;\;\;\;,
\end{equation}
Within the free volume theory, Cohen and Turnbull estimated that the average  free volume goes to zero linearly with the temperature

\begin{equation}\label{vft}
{\left< v_f \right>} = A(T-T_0)   
\;\;\;\;,
\end{equation}

Inserting Eq.\ref{vft}  into Eq.\ref{taufreev} gives a VFT  law for the relaxation time 
\begin{equation}\label{taufreev1}
\frac{\tau}{\tau_0} =  \exp(B/(T-T_0))
\;\;\;\;,
\end{equation}
where A and B  are constants.

Although the relaxation time for many glass formers  can be reasonably well fitted by a VFT law,  the
prediction of a simple exponential for the free volume distribution,  Eq.\ref{pvf}, is not supported
by numerical simulations on hard spheres and experimental data on granular 
materials \cite{Anikeenko08,AsteKGammaPRE08,AsteDeductiveSM} which instead show a very good fit with a Gamma distribution
(see later Eq.\ref{N*2}).

{\bf Cooperative Rearranging Regions}.
An alternative theory was introduced by Adam and Gibbs in 1958 \cite{AG}, who introduced the concept of cooperative rearranging regions. 

The main idea underneath this approach is that, close to the glass transition, due to the crowding of the particles, the decay towards equilibrium of a density fluctuation 
is due to a cooperative rearrangement of an entire  region.  
A cooperative rearranging region of linear size  $\xi$ can be defined as the smallest 
region that can be rearranged without involving particles outside its boundary. 
It is argued that the relaxation time diverges exponentially with the excess entropy with respect to the underline crystalline state $s_c $:

\begin{equation}\label{tauag}
\frac{\tau}{\tau_0} =  \exp(C/(Ts_c))
\;\;\;\;,
\end{equation}
where  C is a constant.
Adam and Gibbs argued that $s_c$ is proportional to  $\xi ^{-d}$, with $d$ the space dimension.
Following Kauzmann suggestion that the excess entropy goes to zero linearly at $T_K$, it follows that the relaxation time diverges according to the  VFT law  Eq.\ref{taufreev} with $T_0=T_K$.

Wolynes and co-workers\cite{kirkpatrick1989scaling}, \cite{xia2000fragilities,lubchenko2007theory} generalized the Adam and Gibbs approach, proposing a mosaic picture, where the 
system is partitioned in droplets of linear size $\xi$. Like in ordinary first order transition the free energy of nucleating a droplet contains a volume term plus a surface term. Within the random first order transition 
the surface term is modified into a term which is proportional to $\xi ^{\theta}$  with $\theta = d/2$.
(see also \cite{bouchaud2004adam} and Berthier and Biroli \cite{berthier2011theoretical} for further elaboration and a discussion of other possible values of $\theta$).  
Consequently the configurational entropy results to be related  to the size of the droplet as $s_c = const/T\xi ^{d-\theta}$ and 
the relaxation time is
\begin{equation}\label{tauw}
\frac{\tau}{\tau_0} =  \exp(\beta \Delta F )
\;\;\;\;,
\end{equation}
where 
$\Delta F$ is the droplet  reconfiguration free energy barrier and it is given by  
\begin{equation}\label{deltaf}
\Delta F \sim (s_c)^{-\theta/(d-\theta)}. 
\;\;\;\;,
\end{equation}
Given that  $\theta = d/2$ the Adam and Gibbs relation,  Eq.\ref{tauag},  is recovered.

{\bf Cell Theory}.
In \cite{AsCo03b,AsCo03,AsCo04} the authors of the present paper, based on lattice theories of liquids\cite{hill1956statistical}, introduced a `cell theory' that combines the ideas of inherent structures, free-volume theory and geometrical packing properties to derive a general theory to understand the complex dynamics of glass-forming liquids, granular packings and amorphous solids.
The main feature of this theory is the demonstration that  thermodynamical properties of these systems can be retrieved from the study of geometrical and topological properties of local configurations only.

In the cell theory, the partition function of a system of $N$ particles in a volume $V$ is:
\begin{equation}\label{FendN1}
Z
=
\sum_{ \{N({\mathbf n})\} }
\Omega(\{N({\mathbf n})\})
e^{ -\beta F(\{ N({\mathbf n}) \}) }
\;\;\;.
\end{equation}
where
\begin{equation}\label{FalphaN1}
F(\{  N({\mathbf n})  \})
=
\sum_{\mathbf n} N({\mathbf n})
\Big\{
\epsilon({\mathbf n})
- kT \big[\ln \frac{v_f({\mathbf n})}{\Lambda^d}  - \ln {\mathcal P}({\mathbf n})\big]
\Big\}
\;\;\;,
\end{equation}
is the  free energy \cite{StillingerandWeber} which is dependent
on the distribution of the cell-shapes
and sizes $\{N({\mathbf
n})\}$ \cite{AsCo03b,AsCo03,AsCo04}; $v_f({\mathbf n})$ is the
\emph{`free volume'} associated with a particle in a cell with a
set of geometrical and topological parameters ${\mathbf n}$;
$\epsilon({\mathbf n})$ is the energy associated with a particle
in a cell with ${\mathbf n}$; ${\mathcal P}({\mathbf n})$ 
is associated with the probability to find  a cell with ${\mathbf
n}$ that is not single-occupied \cite{hill1956statistical}. The quantity $\Omega(\{N({\mathbf n})\})$
in Eq.\ref{FendN1} counts the number of distinct space-partitions
(associated with the inherent states) made with the same set of
$\{N({\mathbf n})\}$. 

The key elements to be estimated in Eq.\ref{FalphaN1} are $\Omega(\{N({\mathbf n})\})$ and ${\mathcal P}({\mathbf n})$.
At high density particles move only locally and all Vorono\"{\i} cells are singly occupied and  ${\mathcal P}({\mathbf n}) \sim 1$.
For what concerns the terms $\Omega(\{N({\mathbf n})\})$,
it was pointed out in
\cite{AsCo04,AsCo03,AsCo03b} that the maximum number of distinct
configurations that -in principle- can be made by positioning in
different ways the $N$ cells distributed in groups of  $N({\mathbf
n})$ is $N! /{ \prod_{\mathbf n}N({\mathbf n})! }$. However, this
maximal value cannot be in general achieved by $\Omega$ since some
of these combinations of cells do not generate space-filling
assemblies and others might be not associated with any inherent
state. As a crude expression of this we can say that only a
fraction of cells $N/\lambda$ can be considered in this exchange
and, analogously, for each kind of cell only the fraction
$N({\mathbf n})/\lambda$ can be exchanged. 
The idea  follows from the mosaic picture of the cooperative rearranging regions  discussed before that the system can be partitioned in a mosaic each region of linear 
size $\xi $  and in each region the effective number of degree of freedom
is proportional to $\xi ^{\theta}$, therefore the total number of degree of 
freedom  $N$  is reduced by a factor 
\begin{equation}\label{xiLa}
\lambda = (\xi/r_0) ^{d-\theta}\;\;,
\end{equation}
with $r_0$ a characteristic size such that $\rho r_0^d =1$, with $\rho$ the  density (the volume fraction).
Under this assumption we have:

\begin{equation}\label{OmXi}
\Omega (\{N({\mathbf n})\})
=
\frac{ (N/\lambda)! }{ \prod_{\mathbf n}(N({\mathbf n})/\lambda)! } \;\;\;.
\end{equation}

From Eqs.\ref{FendN1}, \ref{FalphaN1} and \ref{OmXi}, we obtain
\begin{equation}\label{Zfin1}
Z
=
\sum_{ \{N({\mathbf n})\} }
\exp{
\Big\{
- \beta \sum_{{\mathbf n}} N({\mathbf n})
\big[
\epsilon({\mathbf{n}})
-
kT\big(
\ln{\frac{v_f({\mathbf{n}})}{\Lambda^d {\mathcal P}({\mathbf{n}})}}
-
\frac{1}{\lambda} \ln{\frac{N({\mathbf n})}{N}}
\big)
\big]
\Big\}
}
\;\;\;\;,
\end{equation}
where we used the Stirling approximation: 
\begin{equation}\label{sc1}
\ln\Omega (\{N({\mathbf n})\})
\simeq 
- \sum_{\mathbf n} N({\mathbf n})/\lambda \ln
N({\mathbf n})/N\;\;\;.
\end{equation}

In order to calculate the configurational partition function $Z$
from Eq.\ref{Zfin1} we can introduce a saddle-point approximation
where the sum over all the distributions $\{N({\mathbf n})\}$ is
replaced with the contribution from a distribution $N^*({\mathbf
n})$ which minimizes the free energy. We have:
\begin{equation}\label{lnZ*}
\frac{\ln Z}{N}
=
-\beta \sum_{{\mathbf{n}}} \frac{N^*({\mathbf n})}{N}
\big[
\epsilon({\mathbf{n}})
-
kT\big(
\ln{\frac{ v_f({\mathbf{n}}) }{ \Lambda^d{\mathcal P}({\mathbf{n}}) } }
-
\frac{1}{\lambda} \ln{ \frac{N^*({\mathbf n})}{N} }
\big)
\big] \;\;\;,
\end{equation}
and the distribution $N^*({\mathbf n})$,  that minimizes the system free energy,  is:
\begin{equation}\label{N*1}
N^*({\mathbf n})=
N_0 \left( \frac{v_f({\mathbf{n}})}{{\Lambda^d \mathcal P}({\mathbf{n}})}\right)^{\lambda}
\exp{\big[
-\beta \lambda \epsilon({\mathbf{n}})
-
\mu \lambda v_f({\mathbf{n}})
\big]}
 \;\;\;\;.
\end{equation}
where $N_0$ is a normalization constant and $\mu$ is a Lagrange multiplier associated to the constraint over the total free volume $V_f$.

For simplicity we consider the case of hard spheres\cite{PZ2010} where $\epsilon({\mathbf{n}})=0$. 
However the results obtained in this case can be extended to more general interactions, provided to substitute the free volume, with an effective  temperature dependent free volume which takes into account the Boltzmann weight inside the cell\cite{hill1956statistical}. 
To evaluate the sum in the continuum limit  we must also introduce a function $g(v_f)$ such that $g(v_f)dv_f$ gives the number of cells with free volume between $v_f$ and $v_f+dv_f$, assuming $g(v_f) \sim  v_f^\delta$ then from Eq.\ref{N*1}, the distribution of the cell free-volumes is
\begin{equation}\label{N*2}
p(v_f) = \frac{N^*(v_f)  v_f^\delta}{N}=
\frac{k^{k}}{\Gamma(k)}\frac{v_f^{k-1}}{\left< v_f\right>^{k}}
\exp{\Big( -k \frac{v_f}{\left< v_f \right>} \Big)}
 \;\;\;\;, 
\end{equation}
where $\Gamma(.)$ is the Gamma function, and $k=k_0 + \lambda$  with $k_0=\delta+1$. 
Here ${\lambda}=1$ for low densities. Although the theory is valid for high densities the distribution Eq.\ref{N*2} works well also for small densities. For instance, it was shown in \cite{Pineda04} that, in the zero density limit the Vorono\"{\i} volumes for hard sphere in three dimensions follows Eq.\ref{N*2} with $k=5.586$.

In the case $k=1$  the expression for $p(v_f)$  in Eq.\ref{N*2}  reproduces the expected cell distribution of the free volume theory which is a pure exponential \cite{cohen1959}.
Moreover for $k > 1$,  Eq.\ref{N*2} coincides with the so-called $k$-Gamma distribution \cite{AsteKGammaPRE08} that  has been shown to reproduce  well the distribution of the Vorono\"{\i} volumes in granular packings for the entire range of densities investigated, with $k_0$ being a smooth function of the density.
It was also pointed out in \cite{AsteKGammaPRE08} that 
\begin{equation}\label{kfluctuation}
k = \frac{\left< v_f \right>^2}{\sigma^2}\;\;,
\end{equation}
where $\sigma^2= \left<v^2_f\right> -\left<v_f\right>^2$ is the variance of the free volume.

This equation shows that in the free volume theory $(k=1)$, where there is no cooperative diverging length , the fluctuation of the free volume $\sigma^2$ vanishes as  $\left<v_f\right>^2$, on the contrary if $k$ diverges as $k\sim \left<v_f\right>^{-\nu}$, the  fluctuation of the free volume would vanishes as  $\sigma^2=\left<v_f\right>^{2+\nu}$.  
Since, from Eq.\ref{xiLa}, $k\sim (\xi/r_0) ^{(d-\theta)}$ and the compressibility $\kappa_T\sim \sigma^2$,  it then follows $\kappa_T\sim  (\xi/r_0) ^{-(d-\theta)(1+1/\nu)}$.

Later we will give arguments suggesting that $\nu=1$. Thus if we take the value of $\theta = d/2$ from RFTO, for $d=3$, we have 
\begin{equation}\label{kappaT}
\kappa_T\sim  (\xi/r_0) ^{-9/2}\;\;,
\end{equation}
relating the vanishing of the compressibility to the divergence of the cooperative length.

Recently it has been shown that monodisperse disordered jammed particle are hyperuniform i.e. infinite wavelength volume fraction fluctuation vanish \cite{torquato2003local,zachary2011hyperuniform,hopkins2012nonequilibrium,berthier2011} with a structure factor $S(\mathtt{k})$ that tends to zero linearly in the wavenumber $\mathtt{k}$ as the glassy jammed state is approached, implying the vanishing of the compressibility $\kappa_T\sim S(0)$ and the divergence  of a  length $\xi_{DCF}$, relative to the direct pair correlation function \cite{torquato2003local,zachary2011hyperuniform,hopkins2012nonequilibrium}. 
More precisely since the Fourier transform of the direct pair correlation is given by $c(\mathtt{k}) = (S(\mathtt{k}) -1)/\rho S(\mathtt{k})$.
Close to the jamming glass state we have $c(0) \sim -1/\rho S(0)$, consequently, in  3d, since $\xi_{DCF} \equiv (-c(0))^{1/3}$, we have $\xi_{DCF} \sim (\kappa_T)^{-1/3}$ .  

Interestingly from  Eq.\ref{kappaT}    follows that the cooperative length $\xi$ and the ``hyperuniform'' length $\xi_{DCF}$ are related and, at the jamming glass state, $\xi^{3/2} \sim \xi_{DCF}$. Note that our approach is only valid for equilibrium configurations, however it has been shown that  for hard spheres, even at volume fraction not very close to jamming, the system appears to be out of equilibrium\cite{hopkins2012nonequilibrium}

{\bf Relaxation time}. 
A density fluctuation can relax following two processes. 
The first is a local process  related to the escape from the cage the second is a cooperative process, due to relaxation  of a cooperative rearranging region.  

{\it Escape probability.} 
The escape probability associated with the first relaxation process can be computed by following  the free volume theory \cite{cohen1959,cohen1961,cohen1969} 
\begin{equation}\label{tau}
P_{esc} = \int_{v_f^*}^{\infty} p(v_f)\, dv_f
\;\;\;\;,
\end{equation}

where $p(v_f)$ is given by  Eq.\ref{N*1} and $v_f^*$ is a free volume,
below which the particle cannot escape from the cell.

The integral in Eq.\ref{tau} is:
\begin{equation}\label{tauExact}
 P_{esc}  = \frac{\Gamma({k,k \frac{ v_f^*}{\left< v_f \right>}})}{\Gamma(k)}
\;\;\;\;,
\end{equation}
where $\Gamma(.,.)$ is the upper incomplete gamma function.

{\it Reconfiguration probability.} 
The second mechanism that contributes to the relaxation time is the reconfiguration of a cooperative rearranging region.
The probability that such reconfiguration occurs is given by $P_{crr} \sim 1/\Omega_{\xi}$ where $\Omega_{\xi}$ is 
the total number of configurations inside the droplet \cite{AG}, \cite{xia2000fragilities}. $\ln(\Omega_{\xi}) = \xi^d r_0^{-d} s_c$ 
where $s_c = \frac{1}{N} \ln\Omega$  is the configurational entropy per particle, which from Eq.\ref{sc1} is $s_c = B/\lambda$ with  $B \sim
 -\int_{0}^{\infty} p(v_f) \ln p(v_f)\, d v_f $ being a slowing varying function of the density, which in the following we will consider  constant with values of the order of 1.

In conclusion, since  $\lambda = (\xi/r_0) ^{d-\theta}$,  we have for the configurational entropy 
\begin{equation}\label{sc}
s_c \sim \xi^ {\theta-d}\;\;\;.
\end{equation}
 and for the reconfiguration probability
\begin{equation}\label{pcrr2}
P_{crr} =A \exp(-B(\xi/r_0)^{\theta})\;\;\;.
\end{equation}
where A is a constant. Eq.\ref{pcrr2}  is in agreement with the result of RFOT.

The relaxation times associated with the two mechanisms are respectively proportional to the inverse of the probabilities: $\tau \sim 1/P_{esc}$ and $\tau \sim 1/P_{crr}$.

{\it Relation between $\xi$ and $ {\left< v_f \right>}$}. 
When the free volume per particle  $ {\left< v_f \right>}$ is small, even  smaller than $v_f^*$, a particle can still manage to jump out of the cage albeit with an exponentially small probability, as the free volume distribution exhibits and exponential tail, which allows very rarely a single particle to have a free volume $v_f > v_f^*$.  
In order to have instead a cooperative relaxation of an entire droplet of radius $R$ the free volume $(R /r_0)^{d}   {\left< v_f \right>}$ of  the droplet must be large enough to allow the interior to change from one inherent state to another, keeping the boundary fixed. 
This transition can occur if the available free volume is distributed in such a way that all the free volume is essentially  concentrated in a  region around the surface of volume $R ^{\theta}$, in order to match the internal inherent configuration with the inherent configuration outside the boundary.
The minimum size  $R=\xi$ must be such that the particles in the region near the boundary has just the minimum free volume, roughly of the order of $v^*$, enough to realize the matching. 
This condition corresponds to the marginal stability limit \cite{bouchaud2004adam}; namely, for fixed boundary,  droplets of radius $R<\xi$ are stable, while droplets with radius $R>\xi$ are unstable.
In conclusion this condition implies $(\xi/r_0) ^{d}   {\left< v_f \right>} = k_1 (\xi/r_0) ^{\theta}v_f^*$, where $k_1$ is a constant, from Eq.\ref{xiLa} we obtain:  
\begin{equation}\label{lambda}
\lambda = k_1 v_f^*/{\left< v_f \right>}.
\end{equation}
We expect $k_1 < 1$ because $\lambda$ must become of the order of 1 when the average free volume $\left< v_f \right>$ is larger than $ v_f^*$. 

Using the value $\theta=d/2$
and $(\xi/r_0) ^{d/2} = k_1v_f^*/{\left< v_f \right>}$ it follows  $k=k_0+k_1v_f^*/{\left< v_f \right>}$. 
Note that  (Eq.\ref{lambda}) is valid only for $\lambda>1$.

To calculate the relaxation time we have to distinguish two regimes: i) a low density regime where $\lambda=1$;  ii) an high density regime where $\lambda>1$.

\begin{itemize}
\item[1)] In the low density regime, where $\lambda=1$,  the relaxation time is only given by the escape process. From Eq.\ref{tauExact} we have:
\begin{equation}\label{tau3Exact}
\tau = \frac{\tau_0}{P_{esc}}\;\;,
\end{equation}
with $P_{esc}$ given by Eq.\ref{tauExact} with $k=k_0+1$.

\begin{figure}[t]
\centering
\includegraphics[width= 0.95\columnwidth]{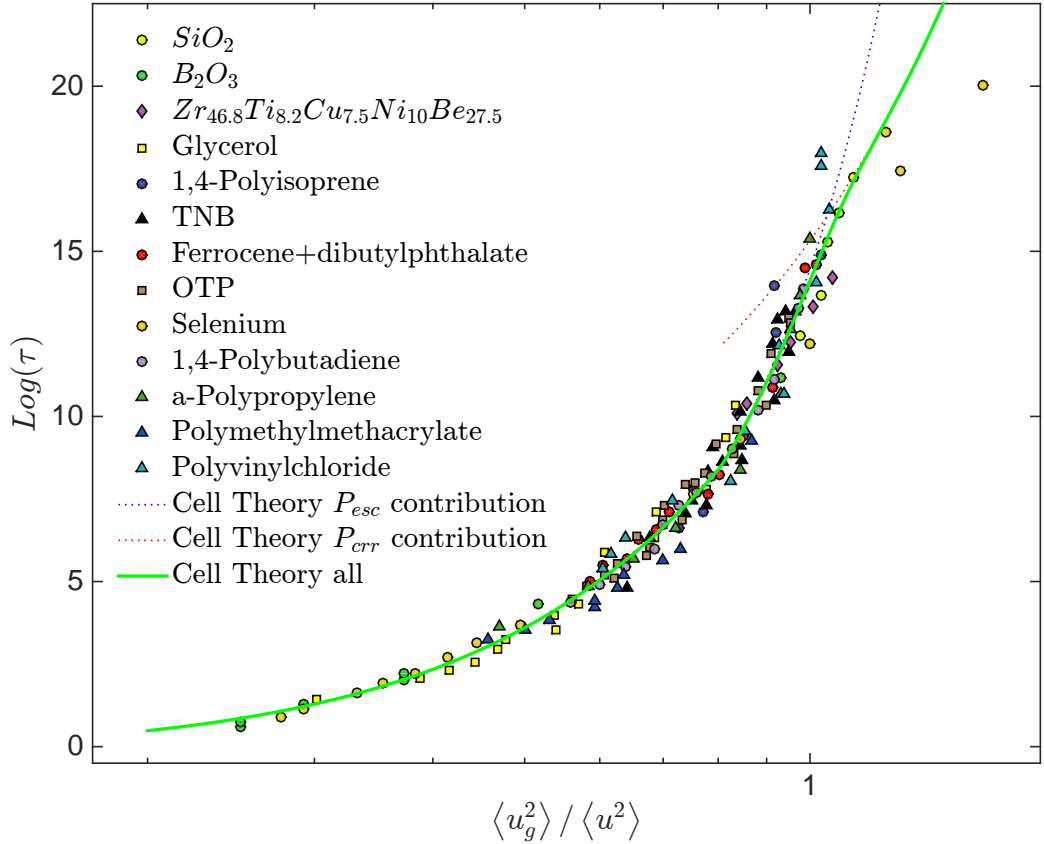}
\caption{\label{figure.fit}
Demonstration that cell theory predicts well experimental measurements of the relaxation  time.
The figure reports the logarithm of the relaxation times $\tau$ vs. relative inverse mean square rattling amplitude $\left< u^2_g \right>/\left< u^2 \right>$ for a large number of different glass-former materials from \cite{Larini} (Fig.3).
The green line is the behaviour given by Eq.\ref{tau4Exact}.
The blue dotted line and the green dotted line are respectively the  contribution from $P_{esc}$ and $P_{crr}$.
One can note that $P_{crr}$ contributes significantly only at very high densities ($\left< u^2_g \right>/\left< u^2 \right> > 1$).
}
\end{figure}

The incomplete  Gamma function can be evaluated for small  and large densities. 
For low densities, when  ${\left< v_f \right>} > v_f^*$, $P_{esc}$ decreases and apparently goes to zero following a power law behaviour:
\begin{equation}\label{powerlaw1}
\frac{\tau}{\tau_0} \sim  ({\left< v_f \right>}/v_f^* - b)^{-\gamma}\;\;\;\;,
\end{equation}
with $b$ close to $1$ and $\gamma =1$. This apparent power law is only a crossover towards 
an exponential behavior for high densities ${\left< v_f \right>} < v_f^*$:
\begin{equation}\label{exponential0}
 \frac{\tau}{\tau_0} \sim  \exp\left(\left(k_0 + 1\right) v^*_f/{\left< v_f \right>}\right)
\end{equation}

\item[2)]  In the density regime, where  $\lambda>1$,  both the escape and collective rearrangement mechanisms contribute to the relaxation time.  
The relaxation time is given by
\begin{equation}\label{tau4Exact}
\frac{\tau}{\tau_0} = 1/(P_{esc} + P_{crr}) 
\;\;\;\;.
\end{equation}
with $P_{esc}$ given by Eq.\ref{tauExact}, $P_{crr}$ given by Eq.\ref{pcrr2} and $k=k_0+k_1v_f^*/{\left< v_f \right>}$. This gives an expression for the relaxation time as function of $v_f^*/{\left< v_f \right>}$. 
\end{itemize}
Using the expression for the incomplete Gamma function evaluated at large  values of $v_f^*/{\left< v_f \right>}$ the relaxation time at high densities is given more explicitly by 
\begin{equation}\label{exponential1}
\frac{\tau}{\tau_0} \sim \frac{1}{\exp\left(-k_0v_f^*/{\left< v_f \right>} - k_1(v_f^*/{\left< v_f \right>})^{2}\right)+  A\exp\left(-k_1Bv^*_f/{\left< v_f \right>}\right)} 
\end{equation}
In a recent work by Larini et al. \cite{Larini} it was pointed out that a large numbers of glass-former materials and several polymers reveal a universal correlation between the structural relaxation time (and the viscosity) and the mean square rattling amplitude (or Debye-Waller  factor), $\left< u^2 \right>$, of the vibrational modes.
It is reasonable to conjecture that  the mean square rattling amplitude must be related with the free volume, with larger amplitudes associated to greater free volume availability.
{\color{black} On the basis of dimensional analysis, we expect $\left< u^2 \right>/\left< u^2_g \right>
=  (\left< v_f \right>/v_g)^{2/3}$, with $ \left< u^2_g \right> $ and $v_g$ respectively the mean square rattling amplitude and the average free volume at glass transition. 
}
By using the experimental data collected by Larini et al. in \cite{Larini} (Fig.3) we can compare the cell theory expression for $\tau$ from Eq.\ref{tau4Exact} with the experimental data. The result is shown in Fig.\ref{figure.fit} were we can see that Eq.\ref{tau4Exact}  reproduces well the experimental findings with a fitting curve (green line) with parameters: $k_0=0.68$, $k_1 = 0.16$, $B = 8.24$, $\tau_0 =  0.86$, $A = 0.018 $ and $v_g = 0.12 v^*_f$.


Finally note that if we use, from the free volume theory, the relation between free volume and temperature
 $\left< v_f \right> \sim T-T_0$  and $\theta = d/2$ we recover the RFOT expression for $\xi \sim (T-T_0)^{-2/d}$
(Eq.\ref{lambda}), $s_c\sim  (T-T_0)$ (Eq.\ref{sc})
and, from Eq.\ref{exponential1}, a VFT for the relaxation time at a temperature very close to the Kauzmann temperature $T_0$.

In conclusions using a cell theory previously developed,  combined with the Adam-Gibbs-Wolyness cooperative rearranging region approach, we have shown that there are two distinct regimes characterizing the relaxation time.
 
In the first regime, where the cooperative length $\xi/r_0=1$, the relaxation time is entirely due a local process Eq.\ref{tau3Exact}. This regime is characterised by an apparent power law, reminiscent of the mode coupling theory behaviour, followed by an exponential behaviour,  typical of the free volume theory. At higher densities the relaxation is due to both escape process (Eq.\ref{tauExact}) and relaxation of cooperative regions (Eq.\ref{pcrr2}). These two processes coexist until at high density the presence of the cooperative regions suppress the escape probability and the only process left is due to the rearranging cooperative regions. For the relaxation due to the local process we use the relation which comes from the free volume theory. This is a simplification as the local relaxation may be more complex due not only to the free volume but to other quantities, involving also second and third neighbors\cite{Leporini}. 

Interestingly our approach relates also the cooperative length $ \xi$  and the hyperuniform uniform length $\xi_{DCF}$ .
 
{\bf Acknowledgements:}
AC would like to thank  A. de Candia, A. Fierro, R. Pastore M. Pica Ciamarra for discussions, and acknowledge financial support from MIUR-FIRB RBFR081IUK and from the CNR-NTU joint laboratory Amorphous materials for energy harvesting applications. 
We also would like to thank D. Leporini, S.Torquato and P. Wolynes for useful comments.


\begin{thebibliography}{29}
\expandafter\ifx\csname natexlab\endcsname\relax\def\natexlab#1{#1}\fi
\expandafter\ifx\csname bibnamefont\endcsname\relax
  \def\bibnamefont#1{#1}\fi
\expandafter\ifx\csname bibfnamefont\endcsname\relax
  \def\bibfnamefont#1{#1}\fi
\expandafter\ifx\csname citenamefont\endcsname\relax
  \def\citenamefont#1{#1}\fi
\expandafter\ifx\csname url\endcsname\relax
  \def\url#1{\texttt{#1}}\fi
\expandafter\ifx\csname urlprefix\endcsname\relax\def\urlprefix{URL }\fi
\providecommand{\bibinfo}[2]{#2}
\providecommand{\eprint}[2][]{\url{#2}}

\bibitem[{\citenamefont{Cohen and Turnbull}(1961)}]{cohen1961}
\bibinfo{author}{\bibfnamefont{M.}~\bibnamefont{Cohen}} \bibnamefont{and}
  \bibinfo{author}{\bibfnamefont{D.}~\bibnamefont{Turnbull}},
  \bibinfo{journal}{The Journ of Chem Phys} \textbf{\bibinfo{volume}{34}},
  \bibinfo{pages}{120} (\bibinfo{year}{1961}).

\bibitem[{\citenamefont{Cohen and Turnbull}(1959)}]{cohen1959}
\bibinfo{author}{\bibfnamefont{M.}~\bibnamefont{Cohen}} \bibnamefont{and}
  \bibinfo{author}{\bibfnamefont{D.}~\bibnamefont{Turnbull}},
  \bibinfo{journal}{The Journ of Chem Phys} \textbf{\bibinfo{volume}{31}},
  \bibinfo{pages}{1164} (\bibinfo{year}{1959}).

\bibitem[{\citenamefont{Cohen and Turnbull}(1969)}]{cohen1969}
\bibinfo{author}{\bibfnamefont{M.}~\bibnamefont{Cohen}} \bibnamefont{and}
  \bibinfo{author}{\bibfnamefont{D.}~\bibnamefont{Turnbull}},
  \bibinfo{journal}{The Journ of Chem Phys} \textbf{\bibinfo{volume}{52}},
  \bibinfo{pages}{30} (\bibinfo{year}{1969}).

\bibitem[{\citenamefont{Adam and Gibbs}(1965)}]{AG}
\bibinfo{author}{\bibfnamefont{G.}~\bibnamefont{Adam}} \bibnamefont{and}
  \bibinfo{author}{\bibfnamefont{J.}~\bibnamefont{Gibbs}}, \bibinfo{journal}{J.
  Chem Physics} \textbf{\bibinfo{volume}{43}}, \bibinfo{pages}{139}
  (\bibinfo{year}{1965}).

\bibitem[{\citenamefont{Kauzmann}(1948)}]{kautz}
\bibinfo{author}{\bibfnamefont{W.}~\bibnamefont{Kauzmann}},
  \bibinfo{journal}{Chem. Rev.} \textbf{\bibinfo{volume}{43}},
  \bibinfo{pages}{219} (\bibinfo{year}{1948}).

\bibitem[{\citenamefont{G\"otze et~al.}(1991)\citenamefont{G\"otze, Hansen,
  Levesque, and Zinn-Justin}}]{gotze1991liquids}
\bibinfo{author}{\bibfnamefont{W.}~\bibnamefont{G\"otze}},
  \bibinfo{author}{\bibfnamefont{J.}~\bibnamefont{Hansen}},
  \bibinfo{author}{\bibfnamefont{D.}~\bibnamefont{Levesque}}, \bibnamefont{and}
  \bibinfo{author}{\bibfnamefont{J.}~\bibnamefont{Zinn-Justin}},
  \emph{\bibinfo{title}{Liquids, freezing and the glass transition}}
  (\bibinfo{year}{1991}).

\bibitem[{\citenamefont{G\"otze}(2008)}]{gotze2009}
\bibinfo{author}{\bibfnamefont{W.}~\bibnamefont{G\"otze}},
  \emph{\bibinfo{title}{Complex Dynamics of Glass-Forming Liquids: A
  Mode-Coupling Theory: A Mode-Coupling Theory}}, vol. \bibinfo{volume}{143}
  (\bibinfo{publisher}{Oxford University Press}, \bibinfo{year}{2008}).

\bibitem[{\citenamefont{Kirkpatrick et~al.}(1989)\citenamefont{Kirkpatrick,
  Thirumalai, and Wolynes}}]{kirkpatrick1989scaling}
\bibinfo{author}{\bibfnamefont{T.}~\bibnamefont{Kirkpatrick}},
  \bibinfo{author}{\bibfnamefont{D.}~\bibnamefont{Thirumalai}},
  \bibnamefont{and} \bibinfo{author}{\bibfnamefont{P.~G.}
  \bibnamefont{Wolynes}}, \bibinfo{journal}{Physical Review A}
  \textbf{\bibinfo{volume}{40}}, \bibinfo{pages}{1045} (\bibinfo{year}{1989}).

\bibitem[{\citenamefont{Xia and Wolynes}(2000)}]{xia2000fragilities}
\bibinfo{author}{\bibfnamefont{X.}~\bibnamefont{Xia}} \bibnamefont{and}
  \bibinfo{author}{\bibfnamefont{P.~G.} \bibnamefont{Wolynes}},
  \bibinfo{journal}{Proceedings of the National Academy of Sciences}
  \textbf{\bibinfo{volume}{97}}, \bibinfo{pages}{2990} (\bibinfo{year}{2000}).

\bibitem[{\citenamefont{Lubchenko and Wolynes}(2007)}]{lubchenko2007theory}
\bibinfo{author}{\bibfnamefont{V.}~\bibnamefont{Lubchenko}} \bibnamefont{and}
  \bibinfo{author}{\bibfnamefont{P.~G.} \bibnamefont{Wolynes}},
  \bibinfo{journal}{Annual Review of Physical Chemistry}
  \textbf{\bibinfo{volume}{58}}, \bibinfo{pages}{235} (\bibinfo{year}{2007}).

\bibitem[{\citenamefont{M{\'e}zard and
  Parisi}(1999)}]{mezard1999thermodynamics}
\bibinfo{author}{\bibfnamefont{M.}~\bibnamefont{M{\'e}zard}} \bibnamefont{and}
  \bibinfo{author}{\bibfnamefont{G.}~\bibnamefont{Parisi}},
  \bibinfo{journal}{Journal of Physics: Condensed Matter}
  \textbf{\bibinfo{volume}{11}}, \bibinfo{pages}{A157} (\bibinfo{year}{1999}).

\bibitem[{\citenamefont{Aste and Coniglio}(2004)}]{AsCo04}
\bibinfo{author}{\bibfnamefont{T.}~\bibnamefont{Aste}} \bibnamefont{and}
  \bibinfo{author}{\bibfnamefont{A.}~\bibnamefont{Coniglio}},
  \bibinfo{journal}{Europhys. Lett.} \textbf{\bibinfo{volume}{67}},
  \bibinfo{pages}{165} (\bibinfo{year}{2004}).

\bibitem[{\citenamefont{Anikeenko et~al.}(2008)\citenamefont{Anikeenko,
  Medvedev, and Aste}}]{Anikeenko08}
\bibinfo{author}{\bibfnamefont{A.~V.} \bibnamefont{Anikeenko}},
  \bibinfo{author}{\bibfnamefont{N.~N.} \bibnamefont{Medvedev}},
  \bibnamefont{and} \bibinfo{author}{\bibfnamefont{T.}~\bibnamefont{Aste}},
  \bibinfo{journal}{Phys. Rev. E} \textbf{\bibinfo{volume}{77}},
  \bibinfo{pages}{031101} (\bibinfo{year}{2008}).

\bibitem[{\citenamefont{Aste and Di{~}Matteo}(2008)}]{AsteKGammaPRE08}
\bibinfo{author}{\bibfnamefont{T.}~\bibnamefont{Aste}} \bibnamefont{and}
  \bibinfo{author}{\bibfnamefont{T.}~\bibnamefont{Di{~}Matteo}},
  \bibinfo{journal}{Phys. Rev. E} \textbf{\bibinfo{volume}{77}},
  \bibinfo{pages}{021309} (\bibinfo{year}{2008}).

\bibitem[{\citenamefont{Aste and Di{~}Matteo}(2007)}]{AsteDeductiveSM}
\bibinfo{author}{\bibfnamefont{T.}~\bibnamefont{Aste}} \bibnamefont{and}
  \bibinfo{author}{\bibfnamefont{T.}~\bibnamefont{Di{~}Matteo}},
  \bibinfo{journal}{arXiv:0711.3239}  (\bibinfo{year}{2007}).

\bibitem[{\citenamefont{Bouchaud and Biroli}(2004)}]{bouchaud2004adam}
\bibinfo{author}{\bibfnamefont{J.-P.} \bibnamefont{Bouchaud}} \bibnamefont{and}
  \bibinfo{author}{\bibfnamefont{G.}~\bibnamefont{Biroli}},
  \bibinfo{journal}{The Journal of chemical physics}
  \textbf{\bibinfo{volume}{121}}, \bibinfo{pages}{7347} (\bibinfo{year}{2004}).

\bibitem[{\citenamefont{Berthier and Biroli}(2011)}]{berthier2011theoretical}
\bibinfo{author}{\bibfnamefont{L.}~\bibnamefont{Berthier}} \bibnamefont{and}
  \bibinfo{author}{\bibfnamefont{G.}~\bibnamefont{Biroli}},
  \bibinfo{journal}{Reviews of Modern Physics} \textbf{\bibinfo{volume}{83}},
  \bibinfo{pages}{587} (\bibinfo{year}{2011}).

\bibitem[{\citenamefont{Aste and Coniglio}(2003{\natexlab{a}})}]{AsCo03b}
\bibinfo{author}{\bibfnamefont{T.}~\bibnamefont{Aste}} \bibnamefont{and}
  \bibinfo{author}{\bibfnamefont{A.}~\bibnamefont{Coniglio}},
  \bibinfo{journal}{Physica A 330} \textbf{\bibinfo{volume}{330}},
  \bibinfo{pages}{189 } (\bibinfo{year}{2003}{\natexlab{a}}).

\bibitem[{\citenamefont{Aste and Coniglio}(2003{\natexlab{b}})}]{AsCo03}
\bibinfo{author}{\bibfnamefont{T.}~\bibnamefont{Aste}} \bibnamefont{and}
  \bibinfo{author}{\bibfnamefont{A.}~\bibnamefont{Coniglio}},
  \bibinfo{journal}{J. Phys.: Condens. Matter} \textbf{\bibinfo{volume}{15}},
  \bibinfo{pages}{S803} (\bibinfo{year}{2003}{\natexlab{b}}).

\bibitem[{\citenamefont{Hill}(1956)}]{hill1956statistical}
\bibinfo{author}{\bibfnamefont{T.}~\bibnamefont{Hill}},
  \emph{\bibinfo{title}{Statistical mechanics}} (\bibinfo{year}{1956}).

\bibitem[{\citenamefont{Stillinger and T.A.Weber}(1982)}]{StillingerandWeber}
\bibinfo{author}{\bibfnamefont{F.}~\bibnamefont{Stillinger}} \bibnamefont{and}
  \bibinfo{author}{\bibnamefont{T.A.Weber}}, \bibinfo{journal}{Phys. Rev. A}
  \textbf{\bibinfo{volume}{25}}, \bibinfo{pages}{978} (\bibinfo{year}{1982}).

\bibitem[{\citenamefont{Parisi and Zamponi}(2010)}]{PZ2010}
\bibinfo{author}{\bibfnamefont{G.}~\bibnamefont{Parisi}} \bibnamefont{and}
  \bibinfo{author}{\bibfnamefont{F.}~\bibnamefont{Zamponi}},
  \bibinfo{journal}{Rev. Mod. Phys.} \textbf{\bibinfo{volume}{82}},
  \bibinfo{pages}{789} (\bibinfo{year}{2010}).

\bibitem[{\citenamefont{Pineda et~al.}(2004)\citenamefont{Pineda, Bruna, and
  Crespo}}]{Pineda04}
\bibinfo{author}{\bibfnamefont{E.}~\bibnamefont{Pineda}},
  \bibinfo{author}{\bibfnamefont{P.}~\bibnamefont{Bruna}}, \bibnamefont{and}
  \bibinfo{author}{\bibfnamefont{D.}~\bibnamefont{Crespo}},
  \bibinfo{journal}{Phys. Rev. E.} \textbf{\bibinfo{volume}{70}},
  \bibinfo{pages}{066119 1} (\bibinfo{year}{2004}).

\bibitem[{\citenamefont{Torquato and Stillinger}(2003)}]{torquato2003local}
\bibinfo{author}{\bibfnamefont{S.}~\bibnamefont{Torquato}} \bibnamefont{and}
  \bibinfo{author}{\bibfnamefont{F.~H.} \bibnamefont{Stillinger}},
  \bibinfo{journal}{Physical Review E} \textbf{\bibinfo{volume}{68}},
  \bibinfo{pages}{041113} (\bibinfo{year}{2003}).

\bibitem[{\citenamefont{Zachary et~al.}(2011)\citenamefont{Zachary, Jiao, and
  Torquato}}]{zachary2011hyperuniform}
\bibinfo{author}{\bibfnamefont{C.~E.} \bibnamefont{Zachary}},
  \bibinfo{author}{\bibfnamefont{Y.}~\bibnamefont{Jiao}}, \bibnamefont{and}
  \bibinfo{author}{\bibfnamefont{S.}~\bibnamefont{Torquato}},
  \bibinfo{journal}{Physical review letters} \textbf{\bibinfo{volume}{106}},
  \bibinfo{pages}{178001} (\bibinfo{year}{2011}).

\bibitem[{\citenamefont{Hopkins et~al.}(2012)\citenamefont{Hopkins, Stillinger,
  and Torquato}}]{hopkins2012nonequilibrium}
\bibinfo{author}{\bibfnamefont{A.~B.} \bibnamefont{Hopkins}},
  \bibinfo{author}{\bibfnamefont{F.~H.} \bibnamefont{Stillinger}},
  \bibnamefont{and} \bibinfo{author}{\bibfnamefont{S.}~\bibnamefont{Torquato}},
  \bibinfo{journal}{Physical Review E} \textbf{\bibinfo{volume}{86}},
  \bibinfo{pages}{021505} (\bibinfo{year}{2012}).

\bibitem[{\citenamefont{Berthier et~al.}(2011)\citenamefont{Berthier,
  Chaudhuri, Coulais, Dauchot, and Sollich}}]{berthier2011}
\bibinfo{author}{\bibfnamefont{L.}~\bibnamefont{Berthier}},
  \bibinfo{author}{\bibfnamefont{P.}~\bibnamefont{Chaudhuri}},
  \bibinfo{author}{\bibfnamefont{C.}~\bibnamefont{Coulais}},
  \bibinfo{author}{\bibfnamefont{O.}~\bibnamefont{Dauchot}}, \bibnamefont{and}
  \bibinfo{author}{\bibfnamefont{P.}~\bibnamefont{Sollich}},
  \bibinfo{journal}{Phys. Rev. Lett.} \textbf{\bibinfo{volume}{106}},
  \bibinfo{pages}{120601} (\bibinfo{year}{2011}).

\bibitem[{\citenamefont{Larini et~al.}(2008)\citenamefont{Larini, Ottochian,
  Michele, and Leporini}}]{Larini}
\bibinfo{author}{\bibfnamefont{L.}~\bibnamefont{Larini}},
  \bibinfo{author}{\bibfnamefont{A.}~\bibnamefont{Ottochian}},
  \bibinfo{author}{\bibfnamefont{C.~D.} \bibnamefont{Michele}},
  \bibnamefont{and} \bibinfo{author}{\bibfnamefont{D.}~\bibnamefont{Leporini}},
  \bibinfo{journal}{Nature Physics} \textbf{\bibinfo{volume}{4}},
  \bibinfo{pages}{42} (\bibinfo{year}{2008}).

\bibitem[{\citenamefont{Puosi and Leporini}(2012)}]{Leporini}
\bibinfo{author}{\bibfnamefont{F.}~\bibnamefont{Puosi}} \bibnamefont{and}
  \bibinfo{author}{\bibfnamefont{D.}~\bibnamefont{Leporini}},
  \bibinfo{journal}{J. Chem. Phys.} \textbf{\bibinfo{volume}{136}},
  \bibinfo{pages}{164901} (\bibinfo{year}{2012}).

\end{thebibliography}

\end{document}